\documentclass[aps,prl,twocolumn,superscriptaddress]{revtex4-2}

\usepackage{graphicx}
\usepackage{dcolumn}
\usepackage{bm}
\usepackage{amsmath}
\usepackage{amssymb}
\usepackage{latexsym}
\usepackage{epsfig}
\usepackage{amsbsy}
\usepackage{array}
\usepackage{amssymb}
\usepackage{setspace}
\usepackage{bm}
\usepackage{epstopdf}
\usepackage{subfigure}
\usepackage{color}
\usepackage{textcomp}

\def\sint{\ifmmode{- \!\!\!\!\!\! \int}
    \else{\hbox{$- \!\!\!\! \int \ $}}\fi}

\begin{document}

\title{The detection of Planck-scale physics facilitated by nonlinear quantum optics}

\author{Wenlin Li}
\email{liwenlin@mail.neu.edu.cn}
\affiliation{College of Sciences, Northeastern University, Shenyang 110819, China}
\author{Chengsong Zhao}
\affiliation{College of Sciences, Northeastern University, Shenyang 110819, China}
\author{Najmeh Eshaqi-Sani}
\affiliation{Department of Mathematical, Physical and Computer Sciences,  University of Parma, I-43100 Parma, Italy}
\author{Zhiyu Jiang}
\affiliation{Center for Quantum Information and Quantum Biology, The University of Osaka, 1-2 Machikaneyama, Toyonaka 560-0043, Japan}
\author{Xingli Li}
\email{xinglili@cuhk.edu.hk}
\affiliation{Department of Physics, The Chinese University of Hong Kong, Shatin, Hong Kong SAR, China}

\date{\today}
\begin{abstract}
A tenet of contemporary physics is that novel physics beyond the Standard Model lurks at a scale related to the Planck length. The development and validation of a unified framework that merges general relativity and quantum physics is contingent upon the observation of Planck-scale physics. Here, we present a fully quantum model for measuring the nonstationary dynamics of a ng-mass mechanical resonator, which will slightly deviate from the predictions of standard quantum mechanics induced by modified commutation relations associated with quantum gravity effects at low-energy scalar. The deformed commutator is quantified by the oscillation frequency deviation, which is amplified by the nonlinear mechanism of the detection field. The measurement resolution is optimized to  a precision level that is $15$ orders of magnitude below the electroweak scale. 
\end{abstract}
\maketitle
At the Planck length $L_p=\sqrt{\hbar G/c^3}\simeq 1.6\times 10^{-35}$\,m, quantum and gravitational effects will become significant simultaneously, giving rise to the emergence of new physics induced by quantum gravitational effects~\cite{Stephen}. It is crucial to observe Planck-scale physics in order to validate competing theoretical frameworks for quantizing gravity, which also posit that the Planck length represents the observational limit of space~\cite{Hossenfelder2013}. Consequently, if a proposed quantum gravity theory degenerates to its corresponding quantum mechanics under a low-energy scalar, it should exhibit a generalized uncertainty principle (GUP) distinct from the Heisenberg uncertainty principle~\cite{Hossenfelder2013,Amati1987}:
\begin{equation}
\begin{split}
\Delta q\Delta p\geq\dfrac{\hbar}{2}\left[1+\beta_0\left(\dfrac{L_p\Delta p}{\hbar}\right)^2\right],
\end{split}
\label{eq:GUP }
\end{equation}
which ensures that space cannot be measured with infinite precision. As a phenomenologically introduced parameter, $\beta_0$ sets a new physical length $\sqrt{\beta_0}L_p$, below which Planckian corrections could become significant and induce new physics. Its value is expected to lie in the range $[1, 10^{34}]$, with the upper bound constrained by the electroweak scale $10^{17}L_p$, beyond which effects would have been detected~\cite{Das2008}.  Nevertheless, this assumption lacks support from first principles, necessitating experimental determination of the upper bound on $\beta_0$. 

The observation and measurement of GUP is based on deviations from standard physics predictions that originate from extended quantum mechanics with a modified canonical commutator $[q,p]_{\beta_0}=i\hbar[1+\beta_0({L_p p}/{\hbar})^2]$. In the vicinity of the electroweak scale, the upper bound of $\beta_0$ are determined to be $\beta_0<10^{36}$ and $\beta_0<10^{34.6}$ from the high-resolution spectroscopy of the ground state Lamb shift~\cite{Das2008,Ali2011} and the 1S-2S level difference~\cite{Quesne2010} of the hydrogen atom, $\beta_0<10^{33.5}$ from the minimum energy of an oscillator measured by AURIGA detector~\cite{Marin2013}, $\beta_0<10^{60}$ from the gravitational waves event~\cite{Das2021}, and $\beta_0<10^{30-41}$ from the decoherence of the neutrino oscillation~\cite{Abi2020}. Recently, the evaluation is optimized by considering the nonstationary dynamics instead of the stationary state. In a classical regime, a non-trivial nonlinearity trajectory corresponding to $\beta_0\simeq 10^{7.4}$ was observed via the measurement of a slight frequency shift of a low-frequency, large-mass resonator ($m\simeq 10^{-2}$\,g)~\cite{Bawaj2015}, and it increases to $\beta_0\simeq 10^{19.3}$ upon measuring a ng-level resonator excited by a piezoelectric ceramic. It is imperative to devise a quantum extension of this scheme that eliminates classical excitation, given that GUP is associated with quantum gravity effects. Nevertheless, the resolution of such a nonstationary scheme is fundamentally constrained by the quality factor ($\text{Q}$-factor) of the resonator, as will become evident in a recent reported fully quantum measurement, which yielded an upper bound $\beta_0<10^{31}$, only three orders of magnitude below the electroweak scale, at which precision the nonlinearity of the microscopic resonator is not observed~\cite{Bonaldi2020}. 

In this letter, we propose a straightforward methodology for gauging GUP based on the dispersive interaction (radiation pressure-like)~\cite{Aspelmeyer2014}.  The entire measurement system, including  the fields that excite and measure the microscopic resonators, is  completely described by \textit{full quantum} theory. Unlike the efforts to enhance the $\text{Q}$-factor of the resonator at the material technology level, the central concept proposed for enhancing the resolution presented is considering a nonlinear responded probe field, whose nonlinear mechanism induces the field to generate high-order sidebands~\cite{Marquardt2006}, which exponentially amplify weak frequency shifts that are originally below the lower bound of resolvable frequencies. The entire scheme is robust and realistic with respect to existing experimental apparatus, in fact, it does not have strict quantitative requirements for any characteristic parameter and does not necessitate precise timing control of the system.


\begin{figure}
\centering
\includegraphics[width=2.85in]{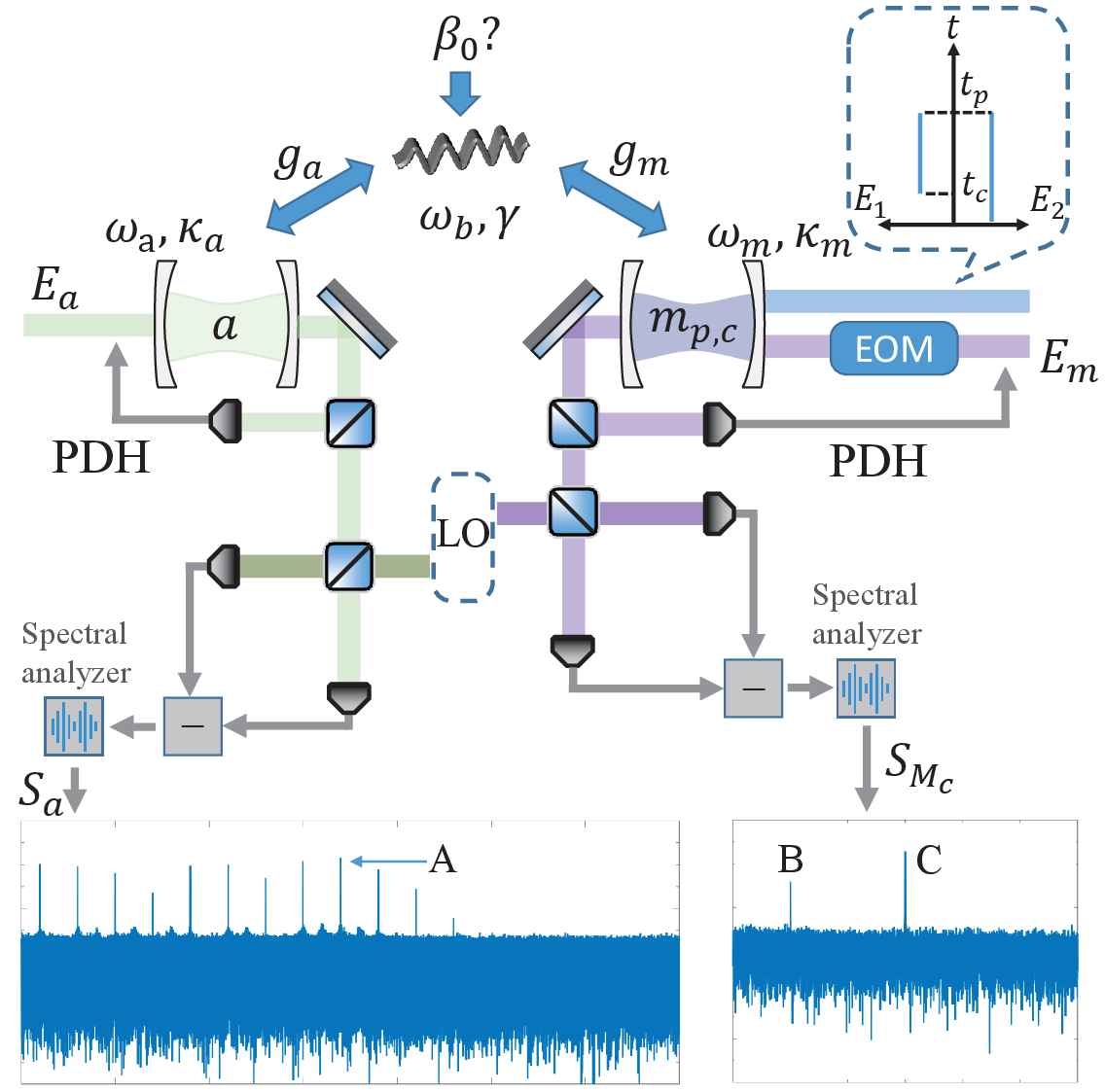}  
\caption{Schematic of the detection scheme: an oscillator modified by GUP that is coupled to two cavity fields through significantly disparate coupling strengths ($g_m\ll g_a$).  The probe field $E_m$ is frequency modulated by an EOM with the parameters $\Omega$ and $\phi$. The dotted box indicates the time sequence of  $E_1$ and $E_2$ which are two components of driving field. The measured spectrum $S_a$ and $S_{M_c}$ are shown at the bottom. 
\label{fig:1}}
\end{figure} 
\textit{GUP-induced frequency modification.}--For a harmonic oscillator with mass $m$ and angular frequency $\omega_b$, the dimensionless Hamiltonian is $H/\hbar=\omega_b(Q^2+P^2)/2$ after defining dimensionless conjugate quantities as coordinate $q=\sqrt{\hbar/(m\omega_b)}Q$ and momentum $p=\sqrt{\hbar m\omega_b}P$,  and the modified canonical commutator is also re-expressed as $[Q,P]_{\beta_{\text{NL}}}=i[1+\beta_{\text{NL}}P^2]$, where $\beta_{\text{NL}}=\beta_0(\hbar m\omega_b/M_p^2c^2)\ll 1$ with  $M_p\simeq 22$~\textmu g is the Plank mass. Priori assumptions of Ref.~\cite{Bawaj2015}  are inherited ensuring the validity of the Heisenberg equations $i\hbar\dot{\hat{s}}=[H,\hat{s}]$ remain unchanged in standard quantum mechanics. Then the commutator can be degenerated into standard form $[Q,\tilde{P}]=i$ by introducing a transform $P=(1+\beta_{\text{NL}}\tilde{P}^2/3)\tilde{P}$~\cite{Bawaj2015,Petruzziello2021}. In this representation, the Hamiltonian of a harmonic oscillator contains an additional correction term induced by the GUP, i.e., $\omega_b\hat{b}^\dagger\hat{b}\rightarrow\omega_b\hat{b}^\dagger\hat{b}+H_{g}/\hbar=\omega \hat{b}^\dagger\hat{b}+\omega_{b}\beta_{\text{NL}}(\hat{b}-\hat{b}^\dagger)^4/{12}$, where $\hat{b}=(Q-i\tilde{P})/\sqrt{2}$ is the annihilation operator. The initial state of the oscillator $\rho_0$ is set to be a Gaussian state with an complex amplitude $\text{Tr}(\rho_0\hat{b})=A_0$, while its dynamics is described by a  differential equation $\dot{{b}}=-i\omega_b {b}+i\beta_{\text{NL}}\omega_b({b}-{b}^*)^3/3$, obtained by replacing the operator $\hat{b}$ in the Heisenberg equation with the complex variable $b$~\cite{Weedbrook2012}. Its solution, $b\simeq A_0e^{-i\\\omega_b(1+\beta_{\text{NL}}\vert A_0\vert^2)t}$, demonstrates an observable effect induced by GUP, i.e., an amplitude-dependent frequency correction in the oscillator frequency. However,  a fundamental limitation exists: the oscillator is inevitably damped by the surrounding environment, resulting in the observable effect induced by GUP existing only in a limited time, that is, it continues until the amplitude decays and tends to zero. The length of the signal in the time domain restricts its resolution in the frequency domain,  given by the minimum distinguishable frequency $\delta\omega\sim \gamma$, which determines observable $\beta_{0}$ has a bound $\beta_{\text{lim}}$ given by:
\begin{equation}
\begin{split}
\beta_{\text{lim}}\sim\dfrac{\gamma M^2_p c^2}{\vert A_0\vert^2\hbar m\omega^2_b}=\dfrac{ M^2_p c^2}{\text{Q}\vert A_0\vert^2\hbar m\omega_b}.
\end{split}
\label{eq:bound}
\end{equation}
The subsequent designed scheme will reduce this measurement limit to $\beta_{\text{lim}}\rightarrow \beta_{\text{lim}}/u$. Furthermore, it ensures that $\beta_0$ will be quantitatively measured if the new physics scale satisfies $\beta_0>\beta_{\text{lim}}$.

\textit{Model and measurement scheme.}--As depicted in Fig.~\ref{fig:1}, the proposed detection device  is comprised of a hybrid system where the mechanical mode, modified by GUP, is coupled to two categories of cavity fields through significantly different coupling strengths (a disparity of two to three magnitudes). This configuration is naturally realized if the two fields possess disparate physical properties (e.g., an optical field and a microwave field) or if the cavity exhibits an extremely asymmetric structure~\cite{Supplemental Material}. The system Hamiltonian including the GUP-induced correction term $H_g$ reads~\cite{Aspelmeyer2014}:
\begin{equation}
\begin{split}
H/\hbar=&\omega_a \hat{a}^\dagger\hat{a}+\omega_m (\hat{m}_p^\dagger\hat{m}_p+\hat{m}_c^\dagger\hat{m}_c)+\omega_{b}\hat{b}^\dagger\hat{b}\\&+[g_a\hat{a}^\dagger\hat{a}-g_m(\hat{m}_p^\dagger\hat{m}_p+\hat{m}_c^\dagger\hat{m}_c)](\hat{b}^\dagger+\hat{b})\\&+(H_d+H_g)/\hbar.
\end{split}
\label{eq:system Hamilton}
\end{equation}
Here $m$ represents the field associated with the weaker field-mechanical coupling strength $g_m$ and it contains two independent modes, distinguished by the subscripts $p$ and $c$ (annihilation operators $\hat{m}_{c,p}$, frequencies $\omega_{m}$, damping rate $\kappa_{m}$). Correspondingly, the field $a$ (annihilation operator $\hat{a}$, frequency $\omega_{a}$, damping rate $\kappa_{a}$) coupled the mechanical mode with a stronger interaction, with the coupling coefficient satisfying $g_a\gg g_m$.  $H_d/\hbar=i\left[E_1(t)e^{-i\omega_1 t}+E_2(t) e^{-i\omega_2 t}\right]\hat{m}_p^\dagger+iE_c\hat{m}_c^\dagger e^{-i\omega_c t-i\phi_0\sin(\Omega_c t)}+iE_a\hat{a}^\dagger e^{-i\omega_A t}+H.c.$ is the Hamiltonian of the coherent drive of the field modes with the corresponding driving frequency $\omega_{1,2,c,A}$  and the drive intensity $E_{1,2,c,a}$~\cite{Mari2009}. The mode $m_c$ is frequency-modulated by a known amount $\phi$ at a frequency $\Omega$ close to the mechanical resonance frequency~\cite{Gorodetksy2010}. The quantitative dynamics analysis is obtained by the stochastic Langevin equations (SLEs), which are valid for Gaussian states~\cite{Weedbrook2012,Wang2014}, and for the field:
\begin{equation}
\begin{split}
\dot{a}=&\left\{i\left[-\Delta'_a-g_a(b^*+b-x_{\text{PDH}})\right]-\kappa_a\right\}a,\\&+E_p+\sqrt{2\kappa_a}a_{in}\\
\dot{m}_p=&\left\{i[-\Delta'_2+g_m(b^*+b-x_{\text{PDH}})]-\kappa_m\right\} m_p\\&+E_1(t)e^{-i\Delta_1t}+E_2(t)+\sqrt{2\kappa_m}m_{p,in},\\
\dot{m}_c=&\left\{i\left[-\Delta'_m+g_m(b^*+b-x_{\text{PDH}})\right]-\kappa_m\right\} m_c\\&+i\phi_0\Omega_c\cos(\Omega_c t)m_c+E_c+\sqrt{2\kappa_m}m_{c,in},
\end{split}
\label{eq:cle equation}
\end{equation}
for the mechanical resonator:
\begin{equation}
\begin{split}
\dot{b}=&\left(-i\omega_{b}-\gamma\right)b+i\omega_{b}\dfrac{\beta_{\text{NL}}}{3} \left(b-b^*\right)^3 +\sqrt{2\gamma}b_{in}\\+&i\left[g_m (\vert m_p\vert^2+\vert m_c\vert^2)-g_a \vert a\vert^2\right],
\end{split}
\label{eq:cle equationb}
\end{equation}
where $\kappa_a$, $\kappa_m$ and $\gamma$ are the decay rates for the field amplitudes and the mechanical amplitude. ${a}_{in}$, ${m}_{p,in}$, ${m}_{c,in}$ and $\hat{b}_{in}$ are the corresponding noise reservoir operators, which are all uncorrelated from each other and can be assumed as usual to be white Gaussian which possess the correlation functions $\langle o(t)_{in}^{*} o(t)'_{in}\rangle=(\bar{n}_o+1/2)\delta(t-t')$ for $o\in\{a,m_p,m_c,b\}$, and $\bar{n}_o=[\exp(\hbar\omega_o/k_bT)-1]^{-1}$ is the mean thermal excitation number for the corresponding mode~\cite{Giovannetti2001}. $\Delta_1=\omega_1-\omega_2$ is the detuning between the two driving frequencies. $\Delta'_{2}=(\omega_{m}-g_{m}x_{\text{PDH}})-\omega_{2}$ [$\Delta'_{c}=(\omega_{m}-g_{m}x_{\text{PDH}})-\omega_{c}$, $\Delta'_{a}=(\omega_{a}+g_{a}x_{\text{PDH}})-\omega_{A}$] is the adjustable detuning between the driving frequency and the corresponding modified cavity frequency instead of the bare cavity frequency, due to the inherent limitations of the PDH cavity lock technology, which can not exclude the correction of the cavity frequency by the low-frequency dynamics of the oscillator $x_{\text{PDH}}(t)=\int_{t-\tau}^td\tau[b(\tau)+b^*(\tau)]/{\tau}$.

The time-dependent driving amplitudes $E_{1}$ and $E_{2}$ serve as pump and cooling fields, respectively, and are modulated according to the sequence
\begin{equation}
   \left\{
    \begin{aligned}
    &E_1(t)=0, E_2(t)=E_2\,\,\,\,(t<t_c) \\
    &E_1(t)=E_1, E_2(t)=E_2\,\,\,\,(t_c<t<t_p) \\
    &E_1(t)=0, E_2(t)=0\,\,\,\,(t>t_p),
    \end{aligned}
    \right.
\end{equation}
with $g_mE_{1,2}\gg g_mE_{m}, g_aE_{a}$, and drive frequencies $\omega_{1,2}$ red-detuned from the cavity resonance. The frequencies $\omega_a$ and $\omega_c$ are tuned to resonate with their respective cavities, such that modes $m_c$ and $a$ function as probe fields. For $t<t_p$, the mean thermal occupancy of the oscillator is suppressed via the anti-Stark effect induced by $E_2$. In the stage $t_c<t<t_p$, interference between the two drive components sustains the mechanical mode in a stable, single-mode coherent state, described by $b(t)=B_0+A(t)e^{-i\Delta_1 t}$, where the slowly varied complex amplitude $A(t)$ obeys the equation formally written as: $\dot{A}\simeq (-i\omega_{\text{eff}}-\gamma_{\text{eff}}) A+\mathcal{F}e^{-i(\Delta_1-\omega_b)t}+\sqrt{2\gamma}b_{in}$. It essentially describes the process of forced vibration of a mechanical mode, where $\omega_{\text{eff}}$ and $\gamma_{\text{eff}}$ denoting the field-renormalized frequency and damping rate, and $\mathcal{F}$ represents the corresponding external force given by:
\begin{equation}
\begin{split}
\mathcal{F}=E_1E_2\sum_{n=-\infty}^{\infty}\dfrac{J^2_n\left[-\xi_m(t)\right]}{[in\omega_b-i\Delta_1-\mathcal{L}_m][-in\omega_b-\mathcal{L}^*_m]},
\end{split}
\label{eq:F2 solution}
\end{equation}
where $\mathcal{L}_m=-i\Delta'_2-\kappa_m$, $\xi_m(t)=2g_m\vert A(t)\vert/\omega_b$ and $J_n$ is the $n$th Bessel function of the first kind. The mechanical mode is pumped to attain an amplitude $\vert A_s\vert^2\simeq {\vert \mathcal{F}\vert^2}/{[(\Delta-\omega_b-\omega_{\text{eff}})^2+\gamma^2_{\text{eff}}]}$ if the stability condition holds in the entire dynamic process. The Bessel function terms in $\mathcal{F}$ captures the saturation effect, whereby further increasing the drive does not significantly enhance the amplitude. While $\gamma_{\text{eff}}$ reflects drive-induced heating, the oscillator can maintain high quantum purity with suitable parameters.

Once the driving field is deactivated after $t_p$, the weak detection fields are insufficient to perturb the oscillator, allowing it to freely evolving as $b(t)\simeq B_0+A(t)e^{-i[\omega_b+\omega_b\beta_{\text{NL}}\vert A(t)\vert^2](t-t_p)}$, with the dissipated amplitude $A(t)=A_se^{-\gamma(t-t_p)}$. $\beta_{\text{NL}}$ is quantitatively measured in this stage, where the mode $a$ records the frequency shift induced by GUP, while mode $m_c$ is used to calibrate the amplitude of the oscillator with $\phi$ and $\Omega$. The output field of modes $m_c$ ($a$) in the time domain is expressed in terms of the corresponding intracavity field and vacuum input noise, i.e., $m_{c,out}=\sqrt{2\kappa_m}m_c-m_{c,in}$ ($a_{out}=\sqrt{2\kappa_a}a-a_{in}$), and it is amplified along the phase direction by a homodyne detection, which obtains $M_{c,out}=K\text{Im}(\sqrt{2\kappa_m}m_c-m_{c,in})$ [$\alpha_{out}=K\text{Im}(\sqrt{2\kappa_a}a-a_{in})$] with a factor $K$ related to the intensity of the local oscillation. In a given time period $[t,t+\Delta t]$, the power spectrum of $O$ ($O\in\{M_{c,out},\alpha_{out}\}$) denoted by $S_O^{t,\Delta t}$, is obtained by: $S_O^{t,\Delta t}(\omega)=\left\vert {(2\pi)}^{-1/2}\int_{t}^{t+\Delta t}dt O(t)e^{-i\omega t}\right\vert^2$ with a upper bound of the frequency resolution $\Delta t^{-1}$. The scale of $\Delta t$ requires $\omega_b^{-1}\ll \Delta t \ll\gamma^{-1}$ thereby ensuring that the modulus of the oscillator is approximately constant over the entire time period, i.e., $\vert A(\forall t'\in[t,t+\Delta t])\vert :=\vert A_t\vert$. Then the spectrum $S^{t,\Delta t}_{O}(\omega)$ can be expressed as:
\begin{equation}
\begin{split}
S_{M_c}^{t,\Delta t}\simeq \dfrac{8K^2E^2_c}{\kappa_m}\left[\xi_{m,t}^2\delta(\omega\pm\omega'_{b,t})+\phi^2\delta(\omega\pm\Omega)\right],
\end{split}
\label{eq:S_m solution}
\end{equation}
and 
\begin{equation}
\begin{split}
S_{\alpha}^{t,\Delta t}&\simeq 2K^2E^2_p\kappa_a\sum_{u=-\infty}^{\infty}[1-(-1)^{u}]\times\\
&\sum_{n=-\infty}^{\infty}\left\vert\left[\dfrac{J_{u-n}\left(\xi_{a,t}\right)J_n\left(\xi_{a,t}\right)}{in\omega'_{b,t}+\kappa_a}\right]\right\vert^2\delta(\omega-u\omega_{b,t}'),
\end{split}
\label{eq:A solution}
\end{equation}
where $\omega'_{b,t}=\omega_b(1+\beta_{\text{NL}}\vert A_t\vert^2)$, $\xi_{m,t}=2g_m\vert A_t\vert/\omega_b$ and $\xi_{a,t}=2g_a\vert A_t\vert/\omega_b$. $g_a$ is selected to be two to three orders of magnitude higher than $g_m$ as aforementioned, enabling $\xi_{m,t}\ll 1$ and $\xi_{a,t}>1$ can be satisfied simultaneously in an appropriate time period.  $\xi_m\ll 1$ implies a negligible nonlinear effect so that $S_{M_c}$ is indeed a linear response spectrum with a double peak structure (shown as `B' and `C' in Fig.~\ref{fig:1}) at both the positive and negative frequencies described by Eq.~\eqref{eq:S_m solution}, which provides a proportional relationship to calibrate the unknown amplitude modulus $A_t$ in experiment:
\begin{equation}
\begin{split}
\text{Estimated} \,\,\vert A_t\vert^2:=\vert A'_t\vert^2=\dfrac{\phi\omega_b}{2g_m}\dfrac{S_{M_c}(\omega'_b)}{S_{M_c}(\Omega_c)}.
\end{split}
\label{eq:A calibrate}
\end{equation}
Correspondingly, $\xi_g>1$ implies that the cavity field $a$ is undergoing significant nonlinear modulation, which indicates its high-order sidebands are excited (e.g. the $17$th sideband `A' in Fig.~\ref{fig:1}). As  illustrated by Eq.~\eqref{eq:A solution}, the position of the $u$-order sideband peak, denote as $\omega_{u,t}$, depends on the corrected oscillator frequency as:
\begin{equation}
\begin{split}
\omega_{u,t}=u\omega_{b,t}'=u\beta_{\text{NL}}\omega_b\vert A_t\vert^2+u\omega_b,
\end{split}
\label{eq:fit}
\end{equation}
 in where the weak correction $\beta_{\text{NL}}$, considered as the slope of the linear relationship between $\omega_{u,t}$ and $\vert A_t\vert^2$, is also amplified by a factor of $u$~\cite{exp1}.
 
\begin{figure}
\centering
\includegraphics[width=2.8in]{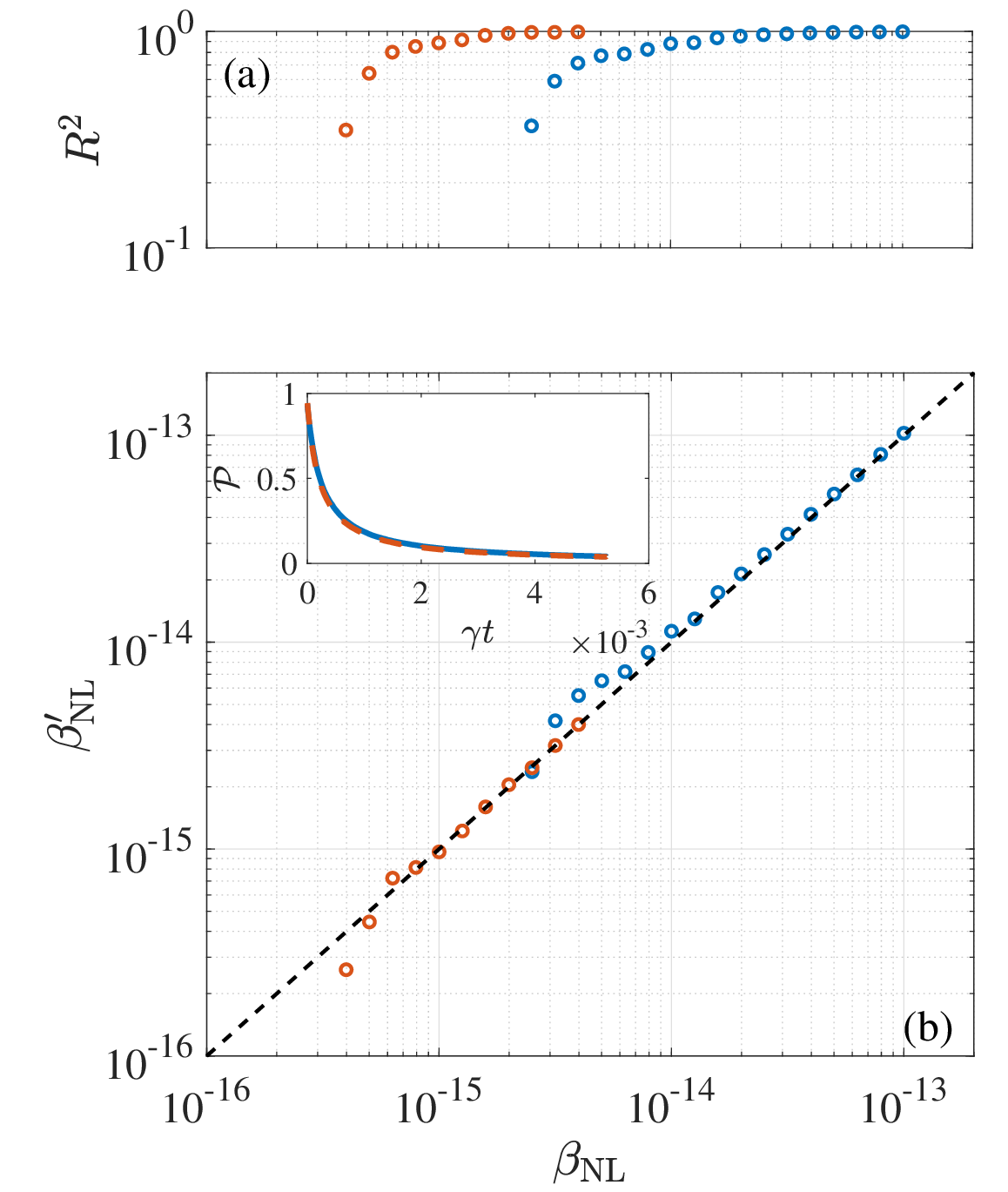}  
\caption{Demonstration of the optimization provided by higher-order sidebands through comparison with results from Ref.~\cite{Li2025}. (a) Linear regression coefficients from the fitting process for each value of $\beta_{\text{NL}}$. (b) Estimated $\beta'_{\text{NL}}$ versus $\beta_{\text{NL}}$. The inset in (b) shows the time evolution of the purity $\mathcal{P}(t)$. Blue points are from data in Ref.~\cite{Li2025}, with our model corresponding to $g_a=0$. Red points are obtained with $g_m/g_a=40$. In addition to the parameters in Tab.~\ref{table:1}, simulations and data processing use $\tau=5.25\times 10^{-6}\gamma^{-1}$, $t_c=0$, $t_p=5.25\times 10^{-4}\gamma^{-1}$, $\Delta t=5.25\times 10^{-4}\gamma^{-1}$, $E_1/\omega_b\simeq 124$, $E_2/\omega_b\simeq 3931$, $E_c/\omega_b\simeq 768$, and $E_p/\omega_b\simeq 38.4$.
\label{fig:2}}
\end{figure}

\textit{Measurements in the quantum regime.}-The efficacy and precision of the measurement apparatus are validated through numerical simulations of the physical implementations shown in Fig.~\ref{fig:1}.  SLEs~\eqref{eq:cle equation} and~\eqref{eq:cle equationb} are simulated with a preset $\beta_{\text{NL}}$ to obtain the dynamical trajectories of the two fields and the resonator. The corresponding output spectra are then computed using input-output relations and spectral analysis; following the aforementioned data-fitting procedure, the estimated value $\beta'_{\text{NL}}$ is extracted. The scheme's generality stems from the nonlinear optical path, featuring strong coupling $g_m$, which acts as an optimized module for integration into existing linear measurement protocols, thereby enhancing ultimate resolution with minimal system perturbation. As an example, we consider a cavity optomechanical system (COM) with parameters listed in Tab.~\ref{table:1}. In the absence of $g_m$, the system reduces to the measurement device of Ref.~\cite{Li2025}, reproducing its results, including a resolution limit of $\beta_{\text{lim}} = 2.5 \times 10^{-15}$ before complete dissipation of the oscillator's purity $\mathcal{P}$. As shown in Fig.~\ref{fig:2}, a coupling $g_m/g_a = 40$ excites the $19$th-order sideband, ensuring the linear regression coefficient $R^2$, as a reliability measure, greater than $0.35$, even in the case of $10^{-15.4}$. This reduces the measurement resolution by at least one order of magnitude while preserving the oscillator's purity, ensuring operation in a quantum regime dominated solely by the zero-point fluctuations of the mechanical resonator.
\begin{table}[]
\caption{Critical parameters in realistic scenarios~\cite{Barzanjeh2015,Heyroth2019, Fan2023,Piergentil2018} \label{table:1}}
\begin{tabular}{c|cccccccc}
\hline
      & $\omega_b/2\pi$&  $\kappa_a/2\pi$ & $\kappa_m/2\pi$   & $\text{Q}$ &   $g_a/2\pi$ &  $g_m/2\pi$ & $m$    \\ 
 Unit& MHz & MHz & MHz  &   &  Hz   & Hz  &ng \\ \hline
COMs& $0.525$ & $2.2$ & $ 2.2$  &  $ 10^9$ &  $ 200$   & $5$  & $50$     \\ 
EOMs& $ 10$ & $ 1$ & $ 2$  &   $10^7$ & $ 115.512$ & $ 0.327$ & $10$  \\
OMMs& $ 40$ & $6$ & $ 4$  &   $10^7$ & $ 8$     & $ 0.2$  & $0.04$         \\    \hline 
\end{tabular}
\end{table}

\begin{figure}
\centering
\includegraphics[width=2.8in]{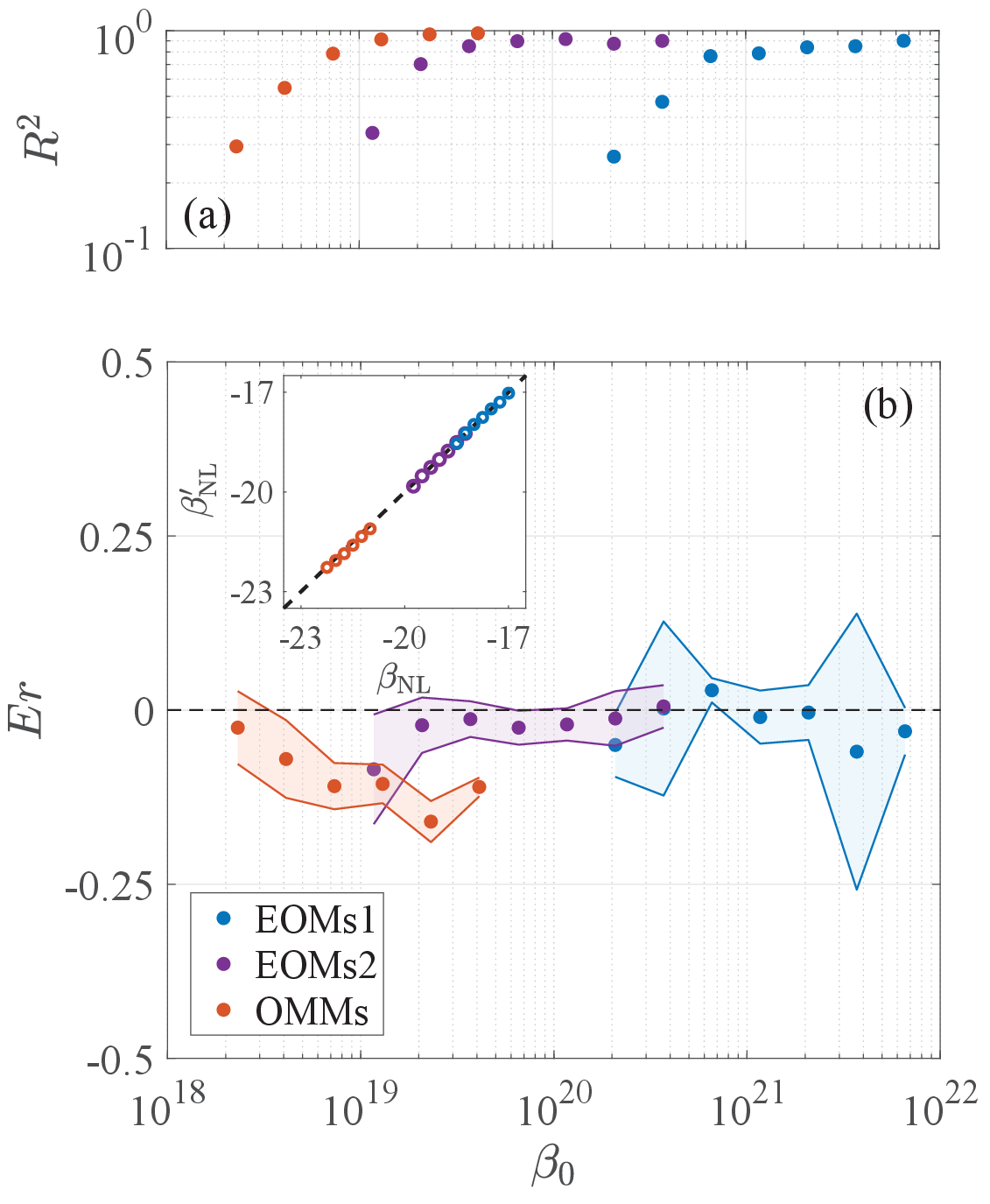}  
\caption{Three sets of simulated measurement results: two of which correspond to EOMs with different parameters (designated as EOMs1 and EOMs2), while the other one corresponds to OMMs. (a): The linear regression coefficient between $\omega_{u,t}$ and $\vert A_t'\vert^2$. (b): The expected value (point) and standard deviation ($68\%$ confidence interval, shaded region) of the measurement error $Er$ obtained from $15$ repetitions. The inset in (b) plots the corresponding average measurement results of the dimensionless parameter $\beta_{\text{NL}}$. In addition to the parameters outlined in Tab.~\ref{table:1}, the remaining parameters utilized in the simulation and data processing are: $\tau=10^{-4}$\,s, $ t_c=10^{-2}$\,s,  $ t_p=0.3$\,s, $\Delta t=0.01$\,s, $E_1/\omega_b\simeq 3.8\times 10^{2} (8.0\times 10^{3})$, $E_2/\omega_b\simeq 1.3\times 10^4$, $E_c/\omega_b=E_p/\omega_b\simeq 8.0\times 10^{-1}$ for EOMs1(2), and  $\tau=2.5\times10^{-5}$\,s, $ t_c=2.5\times 10^{-3}$\,s,  $ t_p=0.075$\,s, $\Delta t=0.0025$\,s $E_1/\omega_b\simeq 3.8\times 10^5 $, $E_2/\omega_b\simeq 1.3\times 10^{6}$, $E_c/\omega_b=E_p/\omega_b\simeq 3.2\times 10^2$ for OMMs.
\label{fig:3}}
\end{figure}

\textit{Scalability to hybrid systems.}-The scheme can be extended to a variety of hybrid systems, leveraging their respective strengths to further improve ultimate resolution.  Here, we consider two such hybrid systems: electro-optomechanics (EOM) comprising a mechanical resonator coupled with microwave and optical cavities, and opto-magnomechanics (OMM) comprising an optical cavity and a yttrium-iron-garnet (YIG) microbridge. The corresponding experimentally achievable parameters are also list in Tab.~\ref{table:1}.  Measurement accuracy and error analysis are quantified by $Er=\log_{10}(\beta'_0/\beta_0)$. The linear regression coefficient is depicted in Fig.~\ref{fig:3}(a), while Fig.~\ref{fig:3}(b) illustrates the expected value and standard deviation ($68\%$ confidence interval) of $Er$, averaged over $15$ repetitions.  It demonstrates that the EOMs offers a more precise assessment of $\beta_0$, while the OMMs provides the highest resolution $\beta_{\text{lim}}\simeq 10^{18.4}$. The enhanced resolution in OMM arises from its ability to achieve stronger driving, yielding a larger oscillator amplitude. Conversely, the enhanced precision in EOM is associated with its weaker coupling ratio $g_a/g_m$. When mode $m$ is in a strong nonlinear regime, EOM provides stronger protection of the linear response of mode $a$, thereby ensuring the accuracy of the calibration relation in Eq.~\eqref{eq:A calibrate}. The inset in (b) shows that for estimating the dimensionless nonlinear parameter $\beta_{\text{NL}}$, the resolution of OMM is approximately $3$ orders of magnitude better than that of EOM. However, this advantage diminishes when evaluating $\beta_0$ due to the smaller mass of the YIG bridge.


\textit{Conclusion and Discussion.}--We have proposed an experiment scheme to measure the tiny frequency shift in nonstationary dynamical oscillation, thereby providing a quantitative evaluation of the GUP in low-energy scalar quantum domain.  The oscillator is prepared to a low-temperature coherent state with a large amplitude and then its free decay process is measured respectively by two quantum probe fields. The corresponding  amplitude is measured by one probe field, which remains linearly responsive, while the frequency shifts are amplified by the high-order sidebands of the other probe field, which exhibits significant nonlinearity. We select experimentally feasible parameters according to three distinct systems. The measurement resolution  was determined to be $\beta_{\text{lim}}=10^{18.4}$ through numerical simulations and $\beta_0$ in GUP will be quantitatively evaluated if it is greater than this limit.

The proposed scheme optimizes the measurement resolution from $\beta_{\text{lim}}\propto \text{Q}^{-1}$ to  $\beta_{\text{lim}}\propto (u\text{Q})^{-1}$ by measuring the $u$th-order sidebands. Thus, $\beta_{\text{lim}}$ can be optimized by inducing the emergence of higher-order sidebands (e.g. increase $E_{1,2}$ or $g_m/g_a$), without significantly enhancing the $\text{Q}$-factor of the oscillator, which is inherently challenging from a materials perspective. Once the \text{Q}-factor has been enhanced, nonlinear measurement will facilitate a more substantial resolution improvement compared to linear systems, as lower $\gamma$ leads to the emergence of higher-order observable sidebands. In comparison with the measurement quantum oscillators reported in Ref.~\cite{Li2025}, the $\beta_{\text{lim}}$ in our scheme is reduced by an order of magnitude while the oscillator is also cooled to near the quantum regime. Additionally, Ref.~\cite{Pikovski2012} theoretically predicts a measurement accuracy optimized to $\beta_{\text{lim}}=10^{12}$ based on strictly symmetric forward-reverse quantum gate operations on the oscillator and repeated quantum measurements. In contrast, our scheme does not involve precise control of the system, which enhances its robustness.

The efficacy of the scheme can be further enhanced by  incorporating additional quantum resources into OMS, such as compressed the oscillator states~\cite{Marti2024}. Finally, our scheme is capable of seamlessly integrating other precision measurements involving small frequency shifts~\cite{He2015} and has the potential to enhance schemes for testing other physical theories, such as observing quantum dynamical effects caused by classical spacetime~\cite{Yang2013}.

We acknowledge Prof. D. Vitali for useful discussions. W. L. is supported by the National Natural Science Foundation of China (Grant No.~12304389), by the Scientific Research Foundation of NEU (Grant No.~01270021920501*115). C. Z. is supported by National Natural Science Foundation of China (Grant No. 12447152). N. E. S. acknowledges financial support from NQSTI within PNRR MUR Project PE0000023-NQSTI. Z. J. is supported by JST, CREST Grant Number JPMJCR24I1, Japan. X. L. is supported by the Guangdong Provincial Quantum Science Strategic Initiative GDZX2404004, and the Space Application System of China Manned Space Program.

\end{document}